\documentclass[aps,twocolumn,nofootinbib,a4paper,superscriptaddress,amsfonts,amssymb,amsmath,floatfix]{revtex4-2}
\pdfoutput=1
\usepackage[utf8]{inputenc}                    
\usepackage[american]{babel}                   
\usepackage[
  colorlinks=true,
  allcolors=teal,
  breaklinks=true
]{hyperref}                                    
\usepackage{microtype}	                       
\usepackage[dvipsnames]{xcolor}                
\usepackage{tikz}
\usetikzlibrary{arrows.meta}
\usetikzlibrary{quantikz2}
\usepackage{physics}
\usepackage{bm}
\usepackage[caption=false]{subfig}

\newcommand{\boxat}[2]{%
  \pgfmathsetmacro{\x}{\xoffset + \radius * cos(#1)}
  \pgfmathsetmacro{\y}{\yoffset + \radius * sin(#1)}
  \pgfmathsetmacro{\rotation}{-90+#1}
  \fill[orange,rotate around={\rotation:(\x,\y)}] (\x-\deltaoffset,\y-\deltaoffset) rectangle (\x+\deltaoffset,\y+\deltaoffset);
  \node[anchor=center,rotate=\rotation] at (\x,\y) {\textbf{#2}};
}
\newcommand{\slotat}[2]{%
  \pgfmathsetmacro{\x}{\xoffset + \radius * cos(#1)}
  \pgfmathsetmacro{\y}{\yoffset + \radius * sin(#1)}
  \pgfmathsetmacro{\rotation}{-90+#1}
  \fill[white,rotate around={\rotation:(\x,\y)}] (\x-\deltaoffset,\y-\deltaoffset) rectangle (\x+\deltaoffset,\y+\deltaoffset);
  \node[orange!70!black,anchor=center,rotate=\rotation] at (\x,\y) {\textbf{#2}};
}
\newcommand{\boxhide}[1]{%
  \pgfmathsetmacro{\x}{\xoffset + \radius * cos(#1)}
  \pgfmathsetmacro{\y}{\yoffset + \radius * sin(#1)}
  \pgfmathsetmacro{\rotation}{-90+#1}
  \fill[gray,opacity=0.8,rotate around={\rotation:(\x,\y)}] (\x-\deltaoffset,\y-\deltaoffset) rectangle (\x+\deltaoffset,\y+\deltaoffset);
}
\newcommand{\connect}[2]{%
  \pgfmathsetmacro{\xstart}{\xoffset + \radius * cos(#1) + \deltaoffset * cos(-90+#1)}
  \pgfmathsetmacro{\ystart}{\yoffset + \radius * sin(#1) + \deltaoffset * sin(-90+#1)}
  \draw[line width=1.0pt,orange!60!black,->] (\xstart,\ystart) arc[start angle=#1-19, end angle=#2+19, radius=\radius];
}

\usepackage{braket}                            
\newcommand{\kket}[1]{\ket{#1}\rangle}         
\newcommand{\bbra}[1]{\langle\bra{#1}}         
\newcommand{\projector}[1]{\ketbra{#1}{#1}}    
\usepackage{dsfont}                            
\newcommand{\id}{\mathds{1}}                   

\usepackage{amsthm}

\theoremstyle{definition}

\begin{document}

\title{No quantum advantage for violating fixed-order inequalities?}

\author{Veronika Baumann}
\thanks{The authors are listed in alphabetical order.}
\affiliation{Institute for Quantum Optics and Quantum Information --- IQOQI Vienna, 1090 Vienna, Austria}
\affiliation{Atominstitut, TU Wien, 1020 Vienna, Austria}

\author{Ämin Baumeler}
\thanks{The authors are listed in alphabetical order}
\thanks{corresponding author, \href{mailto:amin.baumeler@usi.ch}{amin.baumeler@usi.ch}.}
\affiliation{Facolt\`a di scienze informatiche, Universit\`a della Svizzera italiana, 6900 Lugano, Switzerland}
\affiliation{Facolt\`a indipendente di Gandria, 6978 Gandria, Switzerland}

\author{Eleftherios-Ermis Tselentis}
\thanks{The authors are listed in alphabetical order.}
\affiliation{QuIC, Ecole Polytechnique de Bruxelles, C.P. 165, Université Libre de Bruxelles, 1050 Brussels, Belgium}
\affiliation{Faculty of Physics, University of Vienna, 1090 Vienna, Austria}

\begin{abstract}
  \noindent
  In standard quantum theory, the causal relations between operations are fixed.
  One can relax this notion by allowing for dynamical arrangements, where operations may influence the causal relations of future operations, as certified by violation of \emph{fixed-order inequalities,} e.g., the $k$-cycle inequalities.
  Another, non-causal, departure further relaxes these limitations, and is certified by violations of \emph{causal inequalities.}
  In this paper, we explore the interplay between dynamic and indefinite causality.
  We study the $k$-cycle inequalities and show that the quantum switch violates these inequalities without exploiting its indefinite nature.
  We further introduce non-adaptive strategies, which effectively remove the dynamical aspect of any process, and show that the $k$-cycle inequalities become {\em novel causal inequalities;} violating $k$-cycle inequalities under the restriction of non-adaptive strategies requires non-causal setups.
  The quantum switch is known to be incapable of violating causal inequalities, and it is believed that a device-independent certification of its causal indefiniteness requires extended setups incorporating spacelike separation.
  This work reopens the possibility for a device-independent certification of the {\em quantum switch in isolation} via fixed-order inequalities instead of causal inequalities.
  The inequalities we study here, however, turn out to be unsuitable for such a device-independent certification.
  In this work, we initiate the question posed by the title.
  This question, however, remains unanswered.
\end{abstract}

\maketitle

\section{Introduction}
\label{sec:Introduction}

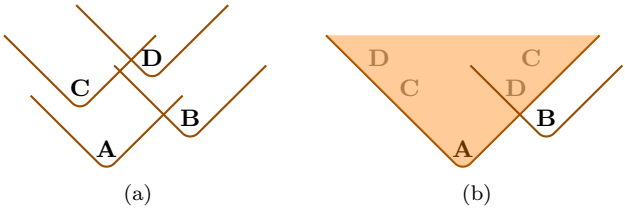
\begin{figure}
  \centering
  \subfloat[\label{subfig:fixedspacetime}]{%
    \begin{tikzpicture}[>={Stealth[scale=0.5]}]
      \foreach \x/\y/\l in {0/0/A, 1.1/0.4/B, -0.35/0.8/C, 0.6/1.2/D} {
        \node at (\x,\y+.3) {\textbf{\l}};
        \pgfmathsetmacro{\lx}{\x-1}
        \pgfmathsetmacro{\ly}{\y+1}
        \pgfmathsetmacro{\rx}{\x+1}
        \draw[thick,orange!60!black,rounded corners] (\lx,\ly) -- (\x,\y) -- (\rx,\ly);
      }
    \end{tikzpicture}
  }
  \qquad
  \subfloat[\label{subfig:dynamicspacetime}]{%
    \begin{tikzpicture}[>={Stealth[scale=0.5]}]
      \foreach \x/\y/\l in {0/0/A, 1.1/0.4/B, -0.35/0.8/C, 0.6/1.2/D} {
        \node[opacity=0] at (\x,\y+.3) {\textbf{\l}};
        \pgfmathsetmacro{\lx}{\x-1}
        \pgfmathsetmacro{\ly}{\y+1}
        \pgfmathsetmacro{\rx}{\x+1}
        \draw[thick,orange!60!black,rounded corners,opacity=0] (\lx,\ly) -- (\x,\y) -- (\rx,\ly);
      }
      \foreach \x/\y/\l/\o in {0/0/A/1, 1.1/0.4/B/1, -.7/.8/C/.5, -1.1/1.2/D/.5, .7/.8/D/.5, .9/1.2/C/.5} {
        \node[opacity=\o] at (\x,\y+.3) {\textbf{\l}};
      }
      \draw[thick,orange!60!black,rounded corners] (-1.8,1.8) -- (0,0) -- (1.8,1.8);
      \fill[orange,rounded corners,opacity=.4] (-1.8,1.8) -- (0,0) -- (1.8,1.8);
      \draw[thick,orange!60!black,rounded corners] (.1,1.4) -- (1.1,0.4) -- (2.1,1.4);
    \end{tikzpicture}
  }
  \caption{%
    In a fixed spacetime (a), the causal relations among the parties are fixed beforehand and independent of their actions.
    (b) In contrast, when the parties are placed on a dynamical spacetime, such as provided by general relativity, then each party's action, e.g., of party $A$, may not only influence the information accessible within the future light cone, but also the causal relations among the parties in the future, i.e., $C, D$.
  }
  \label{fig:spacetime}
\end{figure}

During the last decade, there has been an increasing theoretical and experimental interest in information processing where causality is relaxed, see e.g., Refs.~\cite{%
  chiribella2012,
  colnaghi2012,
  chiribella2013quantum,
  araujo2014,
  feix2015,
  procopio2015experimental,
  pag2016,
  ebler2018,
  quintino2019,
  wei2019experimental,
  caleffi2020,
  felce2020,
  zhao2020,
  taddei2021,
  bavaresco2022,
  quintino2022,
  renner2022,
  abbott2024,
  indefinitecausalexperiments%
}.
Investigations have been heavily based on the process formalism~\cite{ocb2012}, the first level in the hierarchy of higher-order quantum theory~\cite{bisio2019}.
This framework --- motivated by Hardy's~\cite{hardy1, hardy2} groundbreaking attempt to conceptionally reconcile quantum theory with gravity --- is constructed such that individual experiments always conform to the predictions of quantum theory, and without global causal assumptions.
Remarkably, this formalism gives rise to {\em indefinite causal order,} i.e., scenarios where the causal order between events is subject to quantum indefiniteness.
For example, in the quantum switch~\cite{chiribella2013quantum},
Pam, situated in the common causal past of Alice and Bob, coherently controls the causal relation between their experiments.
On the one hand, if Pam prepares the control system in the computational basis, either Alice's experiment can influence Bob's, or {\em vice versa.}
Therefore, the causal relation between Alice and Bob is {\em dynamical,} as opposed to {\em fixed,} see Fig.~\ref{fig:spacetime}.
On the other hand, when Pam prepares the control system in a superposition,
then quantum indefiniteness is injected into the causal relation, and dynamical causal order transforms into {\em indefinite causal order.}

On the level of correlations, this framework allows for violations of causal inequalities~\cite{ocb2012}.
These inequalities serve as device-independent limits to causal explanations.
More concretely, a probability distribution is causal if it can be written as follows
\begin{align}
    p^{\rm causal}(\Vec{a}|\Vec{x})= \sum_k p(k) p_k(a_k|x_k) p_{k,x_k,a_k}(\Vec{a}_{\backslash k}|\Vec{x}_{\backslash k})
\end{align}
where $\Vec{a}=(a_0,a_1 \dots a_{n-1})$ are classical outputs and $\Vec{x}=(x_0,x_1 \dots x_{n-1})$ are classical inputs of parties $i=0,1, \dots n-1$, while $\Vec{a}_{\backslash k}$ and $\Vec{x}_{\backslash k}$ denote the same vectors without the entries with index $k$.
The probabilities~$p_{k,x_k,a_k}(\Vec{a}_{\backslash k}|\Vec{x}_{\backslash k})$, moreover, are required to be causal correlations with {\em one party less.}
Causal correlations appear in scenarios where there exists a partial-order relation $\prec$ according to which each party ($k$) can influence other parties ($j$) in its causal future (if $k \prec j$). This influence also includes the ordering of future parties.
If a~causal inequality is violated, then there is no explanation where present events are caused by past events, and, in turn, act as causes for future events.
This holds even if the causal order is allowed to be {\em dynamically\/} determined, i.e., depend on properties of physical systems.
Processes that violate causal inequalities are called \emph{non-causal.}
The correlations produced by the quantum switch and its generalizations, i.e., processes with quantum control of causal order, however, cannot violate causal inequalities;
they are causal~\cite{oreshkov2016,araujo2015witnessing,ps2021,wechs2021}.
Indefinite causal order present in the quantum switch can, nonetheless, be witnessed device dependently~\cite{araujo2015witnessing,branciard2016}, and even in a semi-device-independent manner~\cite{bavaresco2019semi,dourdent2022}.
Recently, it has been shown that extended setups, where parts of the experiment are performed in spacelike separation,
allow for device-independent certifications of the quantum switch and similar processes~\cite{gppart3,tein2022,netdi2024,vanderlugt2024,chenwangwang2024}.

For some correlations, a fixed-order explanation, i.e., an explanation that assumes a predefined causal structure independent of physical properties, cannot be given.
There, a {\em dynamical\/} determination of the causal order during runtime is necessary.
This necessity is captured through violations of {\em fixed-order inequalities\/}~\cite{isit2014,tselentis2023m}.
Fixed-order correlations are given by probability distributions
\begin{align}
    p^{\rm fix}(\Vec{a}|\Vec{x})=\sum_{\sigma} p(\sigma) \Pi_k p_{k,\sigma}(a_k |a_{\prec_{\sigma} k}, x_{\prec_{\sigma} k}, x_k) \, ,
\end{align}
where the partial order $\sigma$ determines the causal relations of all parties and $\prec_{\sigma} k$ denotes parties in the causal past of $k$ according to $\sigma$. Hence, these correlations describe scenarios where each party can influence other parties in its causal future, but \emph{not} the causal relations among them.

In this paper, we focus on the fixed-order inequality arising from the $k$-cycle game~\cite{tselentis2023m}.
The $k$-cycle game is a communication game played among~$k$ parties, where any number of additional parties may assist these~$k$ parties.
In the presence of one assisting party (Pam), who {\em adaptively\/} prepares the control system, processes akin to the quantum switch are powerful enough to win the~$k$-cycle game.
Also, it is sufficient for Pam to prepare the control system in a classical state; quantum control is not required to perfectly win the game~\cite{tselentis2023m}.

Here, we observe that the causal connections provided by the quantum switch are more than needed to deterministically win the game.
Motivated by this, we study the cyclic quantum switch~\cite{chiribella2021quantum,sazim2021}, and introduce a novel generalization: the sparse quantum switch.
The first allows parties to be causally connected in some total order and all {\em cyclic\/} permutations of it.
The latter --- which is minimal in terms of the causal connections --- only allows {\em a single link\/} to be activated among a cyclic set of channels, while the rest of the parties cannot communicate.

Then we go into a setting where we remove the dynamical component as introduced by Pam's adaptive specification of the control system.
Namely, we constrain Pam to act {\em non-adaptively,} i.e., to output a constant state.
With this, we effectively remove any assisting power of Pam.
We test our {\em restricted\/} setting in relation to the $k$-cycle inequalities, and show that in the absence of any assisting parties, the~$k$-cycle inequalities transform into {\em novel\/} (facet-defining) causal inequalities, the $n$-cycle inequalities, for any  total number $n$ of parties.
In other words, when Pam acts non-adaptively, i.e., for $k=n$, the $n$-cycle inequality, becomes a {\em coinciding facet-defining\/} inequality of causal and fixed-order correlations; violations do not only certify the incompatibility with a static background spacetime, but also with {\em any,} possibly dynamical, causal order.
Finally, we show, invoking non-causal processes, how to deterministically violate the $n$-cycle inequality in the non-adaptive regime.

The investigations carried out here lead us to the following central observation.
In principle, Pam could dynamically specify the causal order among the other parties.
This capability, however, is not inherent to quantum theory, but also present in processes with classical control.
Therefore, if we restrict Pam to non-adaptive strategies, we suppress the classical part, and focus on her quantum capabilities.
Potentially, such quantum capabilities, e.g., specifying the control system in superposition, could outperform any classical non-adaptive strategy.
As we demonstrate below, this is not the case for the~$k$-cycle inequalities studied here.
This fact, which we outline in greater detail, follows from the following.
If in a switch-like process the party in the global future is {\em assisting,} then any output of that party is irrelevant.
But this is tantamount to completely discarding that party.
This, in turn, removes any coherence present in the process: any quantum-over-classical advantage becomes impossible because no ``quantumness'' is left.
Alternatively, if said party in the global future is {\em a playing party,} then coherence may remain intact.
However, as we also mention earlier, this transforms the inequality into a causal inequality, which is known to be unbeatable by switch-like processes.
Although we find no quantum-over-classical advantage for non-adaptively winning the $k$-cycle game,
this possibility remains open
if one considers other fixed-order inequalities~\cite{tselentis2023m}.
The identification of such an advantage would constitute a device-independent certification of switch-like processes {\em in isolation.}

This paper is organized as follows.
In Section \ref{sec:k-cycle games} we present the $k$-cycle game. In Section \ref{sec:Switch processes and k-cycle inequalities} we consider this game within the process formalism where we define adaptive and non-adaptive strategies, and analyze the two party case, i.e., the two-cycle game. We then present its multi-party generalization and the cyclic and sparse quantum switch and show that they deterministically violate the $k$-cycle inequality for {\em any\/} number of parties. In Section \ref{sec:fixed and causal inequalities} we show that if non-adaptive strategies are employed by Pam, then the $n$-cycle inequality, where $n$ is the total number of parties, is also a {\em causal\/} inequality. Then we show how to violate the $n$-cycle inequality with processes beyond coherent control of causal order in Section~\ref{sec:Non-causal processes violating k-cycle inequalities}, and conclude with a series of open questions.

\section{The $k$-cycle game}
\label{sec:k-cycle games}

The causal relations among any set of events on top of a fixed spacetime structure are severely restricted.
When Alice's event, $A$, is in the future of Bob's, $B$, i.e., $B\prec A$, then $B$ cannot be in the future of $A$ ($A\prec B$); the two relations cannot coexist.
This extends to any number of events arranged along a \emph{cycle:}
At most~$k-1$ out of the total number of~$k$ possibilities~$\{i\prec i^+:=i+1 \pmod k\}_{i\in\{0,...,k-1\}}$ may be upheld simultaneously. We take $i=0$ to represent $A$, $i=1$ to represent $B$ etc., and similarly to the cyclic successor $i^+$ of party~$i$, we also define the cyclic predecessor of party $i$ as~$i^{-}:= i-1\pmod k$.
The above limitation can be used as motivation to define the $k$-cycle game~\cite{tselentis2023m}, a communication game that will provide non-trivial bounds under fixed-order constraints. This game features $k$ {\em playing\/} parties, as well as potentially additional {\em non-playing\/} parties who may assist. The total number of parties involved is $n$.
The game is as follows.
A referee picks at random one of the causal relations~$s\prec s^+$, and announces the selection to all parties, playing or not. She then challenges~$s$, who we call ``sender,'' to communicate a random bit $x$ to the cyclic successor~$s^+$, who we call ``receiver.''
The referee provides the bit~$x$ \emph{only} to $s$.
If the causal relations among the parties are {\em fixed,} at least one playing party has no other playing party in its causal past.  If said playing party is picked as the receiver, the communication will fail, and the guess made by the receiver is correct with probability~$1/2$.
So, for any $k$, the winning probability of the $k$-cycle game, if the order between the parties is fixed, is
\begin{equation}\label{eq:cycle_win_fixed}
  p^\text{win}= \frac{1}{2k}\sum_{s,x} p(a=x|s,x)\leq 1-\frac{1}{2k}
    \,,
\end{equation}
where $s$ and $x$ are uniformly distributed, and $a$ is the receiver's guess for the value of $x$.
We call this the $k$-cycle inequality.
Correlations violating this inequality are incompatible with a fixed causal order among the playing parties.

\section{Switch processes and $k$-cycle inequalities}
\label{sec:Switch processes and k-cycle inequalities}
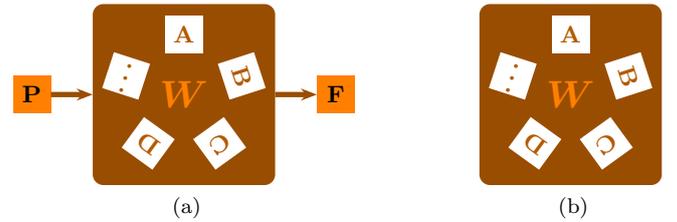
\begin{figure}
  \centering
  \subfloat[\label{fig:new}]{%
    \begin{tikzpicture}[>={Stealth[scale=0.5]}]
      \newcommand{\deltaoffset}{0.25}
      \newcommand{\radius}{0.8}
      \newcommand{\xoffset}{0}
      \newcommand{\yoffset}{0}
      \fill[orange!60!black,rounded corners] (-\radius*1.5,-\radius*1.5) rectangle (\radius*1.5,\radius*1.5);
      \node[orange] at (0,0) {\Large$\bm W$};
      \slotat{90}{A}
      \slotat{18}{B}
      \slotat{-54}{C}
      \slotat{-126}{D}
      \slotat{-198}{\dots}
      \foreach \shift/\l in {-2/P,2/F} {
        \fill[orange] (\shift-\deltaoffset,0-\deltaoffset) rectangle (\shift+\deltaoffset,0+\deltaoffset);
        \node at (\shift,0) {\textbf{\l}};
      }
      \draw[line width=2pt,orange!60!black,->] (-2+\deltaoffset,0) -- ++(+0.55,0);
      \draw[line width=2pt,orange!60!black,<-] (+2-\deltaoffset,0) -- ++(-0.55,0);
    \end{tikzpicture}
  }
  \hfill
  \subfloat[\label{fig:new2}]{%
    \begin{tikzpicture}[>={Stealth[scale=0.5]}]
      \newcommand{\deltaoffset}{0.25}
      \newcommand{\radius}{0.8}
      \newcommand{\xoffset}{0}
      \newcommand{\yoffset}{0}
      \fill[orange!60!black,rounded corners] (-\radius*1.5,-\radius*1.5) rectangle (\radius*1.5,\radius*1.5);
      \node[orange] at (0,0) {\Large$\bm W$};
      \slotat{90}{A}
      \slotat{18}{B}
      \slotat{-54}{C}
      \slotat{-126}{D}
      \slotat{-198}{\dots}
    \end{tikzpicture}
  }
  \caption{%
    We investigate the~$k$-cycle game within the process formalism.
    In case (a), two additional (non-playing) parties,~$P$ and $F$, can influence the game.
    In particular, $P$ and $F$ might employ adaptive strategies where their actions depend on the value of~$s$, which specifies who among the playing parties $A,B,C,D, \dots$ is the sender.
    If they use a non-adaptive strategy, their operations are fixed, and the scenario depicted in (a) is equivalent to the situation shown in (b), where all the parties involved are playing parties. We refer to the game in this scenario as the~$n$-cycle game.
  }
  \label{fig:general}
\end{figure}

Let us now consider the $k$-cycle game within the process formalism, where the \emph{playing parties} $A, B, C,$ etc., as well as the non-playing parties $P$ and $F$ in the global past and future are connected via process $W$, see Fig.~\ref{fig:general}. The parties perform quantum operations, which map a party's quantum input in Hilbert spaces $\mathcal{H}_I$ to their output in Hilbert space $\mathcal{H}_O$. The process $W$ is a generalized quantum supermap~\cite{chiribella2008transforming}, i.e., an operation on quantum operations, and can be represented by the \emph{process matrix,} an operator on the joint space of all the parties' input and output Hilbert spaces.
The probabilities for guess $a$ of the receiver are
\begin{equation}
\label{eq:prob_process}
    p(a|s,x)=
    \Tr\left( W \cdot M_P\bigotimes_{i=0}^{k-1} M_i^{a}(s,x)\otimes M_F \right)\,,
\end{equation}
where $W$ is the process matrix and the $M_i$ are the Choi matrices of the operations performed by the playing parties, which will depend on the value of $s$ and might further depend on the random bit $x$ as well as the receiver party's guess $a$. The Choi matrix $M$ of a completely positive (CP) map $\mathcal{M}:\mathcal{H}_I \rightarrow \mathcal{H}_O$ is defined by
\begin{equation}
    \label{eq:ChoiDef}
    M=  \sum_{i,j} \ketbra{i}{j}_I\otimes \mathcal{M}\big(\ketbra{i}{j}\big)_O
    \,,
\end{equation}
where $\ket{i}_I,\ket{i}_O$ are the computational-basis states of Hilbert spaces $\mathcal{H}_I$ and $\mathcal{H}_O$ respectively. If the map is also trace preserving (TP), the Choi matrix further satisfies
\begin{equation}
  \label{eq:CPTP1}
  \Tr_{O}(M)=\mathds{1}_I
  \,.
\end{equation}
The winning probability for the $k$-cycle game is then given by Eq.~\eqref{eq:cycle_win_fixed},
where the conditional probabilities $p(a|s,x)$ are given by Eq.~\eqref{eq:prob_process}.

Since the parties $P$ and $F$ are not actively playing, meaning they are neither sender nor receiver, and their operations therefore do not depend on $a$ and $x$. However, they do receive $s$, and hence know who are the sender and receiver in each round. In general, the party in the global past is assumed to have a trivial input, i.e., $P$'s operation is the preparation of a quantum state, and the party in the global future is assumed to have trivial output, i.e., $F$'s operation is a measurement. In general, we distinguish between \emph{non-adaptive strategies,} where the operations of $P$ and $F$ do not depend on $s$
\begin{equation}
  \label{eq:PFfixed}
  M_P=\rho_P\,, \quad \text{ and } \quad M^f_F=\Pi^f_F
  \,,
\end{equation}
and \emph{adaptive strategies,} where they do
\begin{equation}
  \label{eq:PFdynamic}
  M_P(s)=\rho(s)_P\,, \quad \text{ and } \quad M^f_F(s)=\Pi^f(s)_F
  \,,
\end{equation}
with $f$ being the outcome of F's measurement. However, as F is a non-playing party, this outcome does not appear in the probabilities relevant for the $k$-cycle game, see Eq.~\eqref{eq:prob_process}. Hence, we effectively always consider
\begin{equation}
  \label{eq:F_trace}
    M_F= \sum_f\Pi^f(s)_F=\mathds{1}_F
  \,,
\end{equation}
regardless of $s$. This means that for the purpose of the $k$-cycle game the action of the party in the global future amounts to just discarding the system they receive.

The sender~$s$ will try to communicate bit $x$ to~$s^+$, but is not required to produce a guess. Using the labeling convention in Sec.~\ref{sec:k-cycle games}, we denote their operation by
\begin{equation}
  \label{eq:sender}
  M^{a}_{i=s}(s,x)= S_{i}(x)
  \,,
\end{equation}
and assume that this \emph{sender operation} is a CPTP map for all $x$. The prototypical example of a sender map is
\begin{equation}
  \label{eq:sender_ex}
  S_{i}(x)= \mathds{1}_{I_i}\otimes \projector{x}_{O_i}
  \,,
\end{equation}
where the sender acts trivially on the input system and encodes their bit $x$ in some basis of their output space.
Conversely, if party $i$ is the receiver, they are required to produce guess $a$. We denote this \emph{receiver operation} by
\begin{equation}
  \label{eq:receiver}
  M^{a}_{i=s^+}(s,x)= R_{i}(a)
  \,,
\end{equation}
which are CP maps such that summing up over all possible guesses $a$ gives a CPTP map, for example
\begin{equation}
  \label{eq:receiver_ex}
  R_{i}(a)= \projector{a}_{I_i}\otimes \frac{1}{2}\id_{O_i}
  \,,
\end{equation}
where the basis $\{\ket{a}\}$ is the one the parties previously agreed upon to encode the bit $x$ in.

First, we investigate quantum processes with classical as well as quantum control of causal order~\cite{chiribella2013quantum,wechs2021,araujo2014,procopio2019communication}, in particular the classical and the quantum switch, and their ability to violate the $k$-cycle inequality, see Eq.~\eqref{eq:cycle_win_fixed}.
The {\em classical $k$-switch\/} is given by
\begin{equation}
   W^k_{\text{CS}}= \sum_{i}\projector{i}_{P_c}\otimes \kket{\pi_i }\bbra{\pi_i}\otimes \projector{i}_{F_c}
   \,,
   \label{eq:cswitch-multi}
\end{equation}
where $\pi_i$ denotes a certain permutation of the $k$ playing parties, and
\begin{equation}
  \begin{split}
    \kket{\pi_i } &= \kket{\mathds{1}}_{P_t I_{\pi_i(0)}} \otimes \kket{\mathds{1}}_{O_{\pi_i(0)}I_{\pi_i(1)}} \otimes \dots \\
    \dots &\otimes \kket{\mathds{1}}_{O_{\pi_i(k-2)}I_{\pi_i(k-1)}}\otimes \kket{\mathds{1}}_{O_{\pi_i(k-1)} F_t}
    \,,
  \end{split}
\end{equation}
with $\kket{\mathds{1}}_{O_\ell I_m}= \sum_i \ket{i}_{O_\ell} \ket{i}_{I_m}$ describing a communication channel from $\ell$ to $m$. The output of $P$ and the input of $F$ consist of both the control, $P_c$ and $F_c$, as well as the target system the playing parties perform their operations on, $P_t$ and $F_t$.
The {\em $k$-party quantum switch\/} is defined as
\begin{equation}
  W^k_{\text{QS}} = \projector{w^k_{\text{QS}}}
  \label{eq:qswitch-multi}
\end{equation}
with
\begin{equation}
  \ket{w^k_{\text{QS}}}= \sum_i\ket{i}_{P_c}\otimes \kket{\pi_i} \otimes \ket{i}_{F_c}
  \,,
  \label{eq:qswitch-multi-ket}
\end{equation}
and represents a process with \emph{coherent control of causal order.} Note that the sums in both Eqs.~\eqref{eq:cswitch-multi} and~\eqref{eq:qswitch-multi-ket} are taken over all the permutations of the $k$-playing parties, and the control system must hence be $k!$-dimensional. For the switch processes we consider \emph{non-adaptive strategies}
\begin{equation}
  \label{eq:PFfixed_switch}
  M_P=\sigma_{P_c}\otimes\rho_{P_t}
  \,,
\end{equation}
and \emph{adaptive strategies}
\begin{equation}
    \label{eq:PFdynamic_switch1}
    M_P(s)=\sigma(s)_{P_c}\otimes\rho_{P_t}\,.
\end{equation}
Since the outcomes of $F$'s measurement are ignored, any off-diagonal elements on the control system do not contribute to the probabilities in Eq.~\eqref{eq:prob_process}.
This means that, for the purpose of the $k$-cycle game, the parties cannot make use of the indefiniteness due to the coherence present in quantum control of causal order.
We now investigate the simplest examples, i.e., $k=2$ and $k=3$, in more detail in the following Secs.~\ref{ssec:The 2-cycle game} and~\ref{ssec:The k-cycle game for 3 or more parties}.

\subsection{The two-cycle inequality}
\label{ssec:The 2-cycle game}

In the two-cycle game, we only have two playing parties, $A$ and B, and if one is the sender the other one is the receiver. For $s=0$, $A$ needs to signal to $B$, while $s=1$ means that $B$ needs to signal to A. We obtain the following probabilities
\begin{align}
\label{eq:prob_2-cycle}
    &p(a|0,x)=
    \Tr\left( W \cdot M_P\otimes S_A(x)\otimes R_B(a)\otimes M_F \right)\,,\\
    &p(a|1,x)=
    \Tr\left( W \cdot M_P\otimes R_A(a)\otimes S_B(x)\otimes M_F \right)\,,
\end{align}
where $W$ is now the two-party version of Eq.~\eqref{eq:cswitch-multi} or~\eqref{eq:qswitch-multi}. For the classical switch $W^2_{\text{CS}}$ and the \emph{non-adaptive strategy} in Eq.~\eqref{eq:PFfixed_switch}, we find that
\begin{equation}
  p^{\rm win}\leq \frac{3}{4}
  \,,
\end{equation}
regardless of which state $P$ prepares.
This is due to the fact that, if $P$ emits a constant state on the control, the process $W$ is a mixture of processes with fixed causal order.
Since, by definition, the latter obey the $k$-cycle inequality, so does the classical switch.
This is true for any number of playing parties, meaning that for non-adaptive strategies and processes defined by Eq.~\eqref{eq:cswitch-multi}, the $k$-cycle inequality is always satisfied.
However, for the adaptive strategy in Eq.~\eqref{eq:PFdynamic_switch1}, we obtain a winning probability of
\begin{equation}
  p^{\rm win}\leq 1
  \,,
\end{equation}
and show that this upper bound can indeed be reached. This means that the $k$-cycle inequality can be violated up to the maximal value.
The detailed calculations can be found in App.~\ref{app:The classical 2-switch}. We now present a specific adaptive strategy that gives this maximal violation.

The non-playing party $P$ can encode the bit $s$ in the control in order to activate the respective communication channel, as well as provide an arbitrary input state of the target system
\begin{equation}
  M_P(s)=\projector{s}_{P_c}\otimes\rho_{P_t}
  \,.
  \label{eq:MP}
\end{equation}
Let the sender and receiver operations furthermore be given by Eqs.~\eqref{eq:sender_ex} and~\eqref{eq:receiver_ex} respectively. This means we have
\begin{align}
    p(a|0,x)&= \bbra{\pi_0}\rho_{P_t}\otimes \projector{x}_{O_A} \otimes \projector{a}_{I_B}\kket{\pi_0} \\
    &= \Tr(\rho)\delta_{ax}\,, \\
    p(a|1,x)&= \bbra{\pi_1}\rho_{P_t}\otimes \projector{a}_{I_A} \otimes \projector{x}_{O_B} \kket{\pi_1}\\
    &= \Tr(\rho)\delta_{ax}\,,
\end{align}
where we left out identity matrices to shorten the notation. This then gives the winning probability of
\begin{equation}\label{eq: violation classical switch}
    p^{\rm win}=\frac{1}{4}\left(\sum_x p(x|0,x)+p(x|1,x)\right) = 1\,.
\end{equation}
Note that this winning strategy is the quantum version of a similar but purely classical strategy discussed in Ref.~\cite{tselentis2023m}.

The quantum switch contains the classical switch
\begin{align}
  W^2_{\text{QS}}&= \sum_{ij}\ketbra{i}{j}_{P_c}\otimes \kket{\pi_i} \bbra{\pi_j}\otimes \ketbra{i}{j}_{F_c} \\
   &= W^2_{\text{CS}} +\ketbra{0}{1}_{P_c}\otimes \kket{\pi_0} \bbra{\pi_1}\otimes \ketbra{0}{1}_{F_c} \\
   &\quad+ \ketbra{1}{0}_{P_c}\otimes \kket{\pi_1} \bbra{\pi_0}\otimes \ketbra{1}{0}_{F_c} \nonumber
   \,,
\end{align}
and hence, we can again use the adaptive winning strategy above to maximally violate the two-cycle inequality. Note that this is true for any number of parties, since the process defined by Eq.~\eqref{eq:cswitch-multi} can be obtained from Eq.~\eqref{eq:qswitch-multi} by deleting the terms containing off-diagonal elements on $P_c$ and $F_c$. As mentioned before, in the $k$-cycle game, the parties cannot utilize the ``cross terms'' distinguishing the quantum switch from its classical counterpart. Therefore, quantum-switch processes in general do not lead to violations of the $k$-cycle inequality for \emph{non-adaptive strategies}. There is no quantum-over-classical advantage for switch-like processes with control of causal order in violating the~$k$-cycle inequality.

One might want to consider scenarios where the causal indefiniteness of quantum-switch processes is used in the game. A straightforward way of doing so is to include $F$'s outcome $f$ in the game by making $F$ a playing party. Note, however, that this automatically means that the $k$-cycle inequality is satisfied:
By construction, $F$ cannot communicate to any other playing party giving, from which we get a contribution of one half to $p^{\rm win}$ whenever $F$ is chosen to be the sender.
Alternatively, one can look for modified versions of the $k$-cycle game that incorporate the outcome $f$, as well as other fixed-order inequalities, see Ref.~\cite{tselentis2023m}.

\subsection{The $k$-cycle inequality for multiple parties}
\label{ssec:The k-cycle game for 3 or more parties}

We now discuss how the winning strategy for the two-cycle game in the previous section readily generalizes to more parties.
This means that both the classical and quantum-switch processes in Eqs.~\eqref{eq:cswitch-multi} and \eqref{eq:qswitch-multi} can maximally violate the $k$-cycle inequality for an \emph{adaptive strategy} of the non-playing party $P$. For three or more playing parties in the $k$-cycle game, we have parties that are neither sender nor receiver in each round. These parties implement the ``do nothing'' operation defined by the Choi matrix
\begin{equation}
  M_{i}^{\perp}= \mathds{1}_{I_i}\otimes \frac{1}{2}\mathds{1}_{O_i}
  \,.
  \label{eq:no_s no_r}
\end{equation}
Importantly, for three or more playing parties, the processes in Eqs.~\eqref{eq:cswitch-multi} and \eqref{eq:qswitch-multi} contain many more ways for $A, B, C,$ etc.\ to signal to each other than are necessary for winning the $k$-cycle game. Already in the case of three playing parties, $A, B, C,$ the control system is six-dimensional while there are only three choices for sender and receiver. Let us label the permutations as follows
\begin{align}
  \pi_0&= A B C \label{eq:pi0}\\
  \pi_1&= B C A \label{eq:pi1}\\
  \pi_2&= C A B \label{eq:pi2}\\
  \pi_3&= A C B \label{eq:pi3}\\
  \pi_4&= B A C \label{eq:pi4}\\
  \pi_5&= C B A \label{eq:pi5}\,.
\end{align}
That way we ensure that the adaptive strategy given by Eq.~\eqref{eq:MP} with $s\in \{0,1,2\}$ allow for winning the $k$-cycle game with certainty if the operations of sender and receiver are given by Eqs.~\eqref{eq:sender_ex} and~\eqref{eq:receiver_ex}. Analogous to the two-cycle game, we find that
\begin{align}
   p(a|s,x)&=
    \bbra{\pi_s}\rho_{P_t}\otimes M_{s^{-}}^{\perp} \otimes S_{s}(x)\otimes R_{s^{+}}(a) \kket{\pi_s} \\
    &= \Tr(\rho)\delta_{ax}
\end{align}
for all values of $s$, which gives the winning probability
\begin{equation}
  p^{\rm win}=\frac{1}{6} \sum_{s,x} p(x|s,x) =1\,,
  \label{eq:winprob3-max}
\end{equation}
for both $W^3_{\text{CS}}$ and $W^3_{\text{QS}}$, since we have, just like in the two-cycle game, used the quantum switch as a classical switch. In this winning strategy we have not used the terms of the switch processes with control values $i\in\{3,4,5\}$. As mentioned before, for \emph{any}~$k> 2$, the quantum switch allows for non-necessary possibilities, i.e., $k!$ orders out of which only $k$ are required for the $k$-cycle game. Motivated by this observation, we propose two generalizations of the quantum switch that are tailored to the~$k$-cycle game.

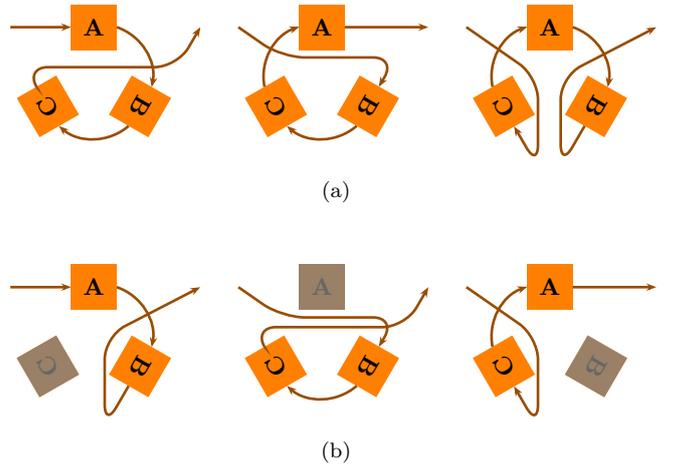
\begin{figure}
  \centering
  \subfloat[\label{fig:cqs}]{%
    \begin{tikzpicture}[>={Stealth[scale=0.5]}]
      \newcommand{\deltaoffset}{0.3}
      \newcommand{\radius}{0.7}
      \newcommand{\xoffset}{0}
      \newcommand{\yoffset}{0}
      \boxat{90}{A}
      \boxat{330}{B}
      \boxat{210}{C}
      \connect{90}{-30}
      \connect{330}{210}
      \draw[line width=1.0pt,orange!60!black,->] (\xoffset-\radius-0.4,\yoffset+\radius) -- ++(0.8,0);
      \pgfmathsetmacro{\x}{\xoffset + \radius * cos(210) + \deltaoffset * cos(-90+210)}
      \pgfmathsetmacro{\y}{\yoffset + \radius * sin(210) + \deltaoffset * sin(-90+210)}
      \draw[line width=1.0pt,orange!60!black,->,rounded corners=3mm] (\x,\y) -- ++(-90+210:.3) -- ++(2,0) -- (\xoffset+2*\radius,\yoffset+\radius);
      \renewcommand{\xoffset}{3}
      \renewcommand{\yoffset}{0}
      \boxat{90}{A}
      \boxat{330}{B}
      \boxat{210}{C}
      \connect{330}{210}
      \connect{210}{90}
      \draw[line width=1.0pt,orange!60!black,->,rounded corners=3mm] (\xoffset-\radius-0.4,\yoffset+\radius) -- ++(-35:.7) -- ++(1.5,0) -- +(330-90:.45);
      \pgfmathsetmacro{\x}{\xoffset + \radius * cos(90) + \deltaoffset * cos(-90+90)}
      \pgfmathsetmacro{\y}{\yoffset + \radius * sin(90) + \deltaoffset * sin(-90+90)}
      \draw[line width=1.0pt,orange!60!black,->,rounded corners=3mm] (\x,\y) -- ++(-90+90:.6) -- (\xoffset+2*\radius,\yoffset+\radius);
      \renewcommand{\xoffset}{6}
      \renewcommand{\yoffset}{0}
      \boxat{90}{A}
      \boxat{330}{B}
      \boxat{210}{C}
      \connect{210}{90}
      \connect{90}{-30}
      \draw[line width=1.0pt,orange!60!black,->,rounded corners=3mm] (\xoffset-\radius-0.4,\yoffset+\radius) -- ++(-35:1.15) -- ++(0,-1.2) -- +(210-90:0.63);
      \pgfmathsetmacro{\x}{\xoffset + \radius * cos(330) + \deltaoffset * cos(-90+330)}
      \pgfmathsetmacro{\y}{\yoffset + \radius * sin(330) + \deltaoffset * sin(-90+330)}
      \draw[line width=1.0pt,orange!60!black,->,rounded corners=3mm] (\x,\y) -- ++(-90+330:0.63) -- ++(0,1.2) -- (\xoffset+2*\radius,\yoffset+\radius);
    \end{tikzpicture}
  }

  \vspace{3ex}
  \subfloat[\label{fig:sqs}]{%
    \begin{tikzpicture}[>={Stealth[scale=0.5]}]
      \newcommand{\deltaoffset}{0.3}
      \newcommand{\radius}{0.7}
      \newcommand{\xoffset}{0}
      \newcommand{\yoffset}{0}
      \boxat{90}{A}
      \boxat{330}{B}
      \boxat{210}{C}
      \boxhide{210}
      \connect{90}{-30}
      \draw[line width=1.0pt,orange!60!black,->] (\xoffset-\radius-0.4,\yoffset+\radius) -- ++(0.8,0);
      \pgfmathsetmacro{\x}{\xoffset + \radius * cos(330) + \deltaoffset * cos(-90+330)}
      \pgfmathsetmacro{\y}{\yoffset + \radius * sin(330) + \deltaoffset * sin(-90+330)}
      \draw[line width=1.0pt,orange!60!black,->,rounded corners=3mm] (\x,\y) -- ++(-90+330:0.63) -- ++(0,1.2) -- (\xoffset+2*\radius,\yoffset+\radius);
      \renewcommand{\xoffset}{3}
      \renewcommand{\yoffset}{0}
      \boxat{90}{A}
      \boxat{330}{B}
      \boxat{210}{C}
      \boxhide{90}
      \connect{330}{210}
      \draw[line width=1.0pt,orange!60!black,->,rounded corners=3mm] (\xoffset-\radius-0.4,\yoffset+\radius) -- ++(-35:.7) -- ++(1.5,0) -- +(330-90:.45);
      \pgfmathsetmacro{\x}{\xoffset + \radius * cos(210) + \deltaoffset * cos(-90+210)}
      \pgfmathsetmacro{\y}{\yoffset + \radius * sin(210) + \deltaoffset * sin(-90+210)}
      \draw[line width=1.0pt,orange!60!black,->,rounded corners=3mm] (\x,\y) -- ++(-90+210:.3) -- ++(2,0) -- (\xoffset+2*\radius,\yoffset+\radius);
      \renewcommand{\xoffset}{6}
      \renewcommand{\yoffset}{0}
      \boxat{90}{A}
      \boxat{330}{B}
      \boxat{210}{C}
      \boxhide{330}
      \connect{210}{90}
      \draw[line width=1.0pt,orange!60!black,->,rounded corners=3mm] (\xoffset-\radius-0.4,\yoffset+\radius) -- ++(-35:1.15) -- ++(0,-1.2) -- +(210-90:0.63);
      \pgfmathsetmacro{\x}{\xoffset + \radius * cos(90) + \deltaoffset * cos(-90+90)}
      \pgfmathsetmacro{\y}{\yoffset + \radius * sin(90) + \deltaoffset * sin(-90+90)}
      \draw[line width=1.0pt,orange!60!black,->,rounded corners=3mm] (\x,\y) -- ++(-90+90:.6) -- (\xoffset+2*\radius,\yoffset+\radius);
    \end{tikzpicture}
  }
  \caption{%
    (a)
    In the cyclic quantum switch, the target system traverses in sequence all parties in a cyclic order.
    The control system specifies the initial party.
    (b)
    The sparse quantum switch:
    Depending on the value of the control system, a {\em single\/} link between two parties is activated.
    All remaining parties are causally disconnected.
  }
  \label{fig:qs}
\end{figure}

\subsubsection{Cyclic quantum switch}
\label{sssec:Cyclic quantum switch}
We present a first known simplified version of the quantum switch where not all permutations of the parties are attainable, but only cyclic ones are, see Refs.~\cite{chiribella2021quantum,sazim2021} and Fig.~\ref{fig:cqs}.
In the three-party case, for instance, the cyclic quantum switch effectively cancels the permutations~\eqref{eq:pi3}--\eqref{eq:pi5} present in the three-party quantum switch.
Such a cancellation of permutations only occurs for three or more parties.
For two parties, both processes coincide.

The cyclic quantum switch operates as follows.
If the control system is in the state~$\ket i$, then the target system initially traverses party~$i$, then~$i^+$, etc., until it reaches $i$'s cyclic predecessor~$i^-$.
From there, the target system is passed to $F$.
As a process matrix, the cyclic quantum switch for~$k$ parties is
\begin{equation}
  W_\text{CQS}^k = \projector{\omega^k}
  \,,
\end{equation}
with
\begin{equation}
  \ket{\omega^k}
  =
  \sum_{i}
  \ket{i}_{P_c} \otimes
  \kket{\pi^{ \circlearrowright}_i} \otimes \ket{i}_{F_c}
  \,,
\end{equation}
where~$\pi^{ \circlearrowright}_i$ is the~$i$th cyclic permutation of the labels~$A,B,\dots$:~$\pi^{ \circlearrowright}_i(j) = i+j \pmod k$.

\subsubsection{Sparse quantum switch}
\label{sssec:Sparse quantum switch}
As per definition of the~$k$-cycle game, for each choice of~$s$, not all parties are required to exchange information in order to win the game.
Instead, only party~$s$ needs to convey the bit~$x$ to party~$s^+$.
This motivates a further simplification:
The definition of a sparse quantum switch where a {\em single\/} link between two successive parties is activated, and all other parties are not communicating, see Fig.~\ref{fig:sqs} and App.~\ref{app:proof} for a proof of correctness.
Here, if the control system is in the state~$\ket{i}$, then that said link is activated: The target system evolves from $P$ to party~$i$, then to party~$i^+$, and finally to $F$.
Since in any case all parties receive a quantum system upon which they may act, we introduce for each party~$j$ a dummy system~$D_j$.
These dummy systems are prepared in the past by $P$ and finally reach $F$ in the future; these systems {\em cannot\/} be used for interparty communication.
The sparse quantum switch for~$k$ parties is
\begin{equation}
  W_\text{SQS}^k = \projector{\Sigma^k}
  \,,
  \label{eq:sqs}
\end{equation}
with
\begin{equation}
  \ket{\Sigma^k}
  =
  \sum_{i\in[k]}
  \ket{i}_{P_c} \otimes
  \kket{\lambda_i} \otimes
  \kket{\Delta_i}  \otimes \ket{i}_{F_c}
  \,.
\end{equation}
Here,~$\kket{\lambda_i}$ establishes the desired communication link,
\begin{equation}
  \kket{\lambda_i}
  =
  \kket{\id}_{P_t,I_{i}} \otimes
  \kket{\id}_{O_{i},I_{i^{+}}} \otimes
  \kket{\id}_{O_{i^{+}},F_t}\,,
\end{equation}
and~$\ket{\Delta_i}$ provides the dummy systems:
\begin{equation}
  \begin{split}
    \kket{\Delta_i}
    &=
    \kket{\id}_{P_{D_i},F_{D_i}} \otimes
    \kket{\id}_{P_{D_{i^{+}}},F_{D_{i^{+}}}}
    \\
    &\qquad
    \bigotimes_{j\in[k]\setminus\{i,i^{+}\}}
    \kket{\id}_{P_{D_j},I_j} \otimes
    \kket{\id}_{O_j, F_{D_j}}
    \,.
  \end{split}
\end{equation}
Note that in~$\kket{\Delta_i}$, the dummy systems for the parties~$i$ and~$i^+$ bypass the respective parties, and in contrast are directly forwarded from $P$ to $F$.
The reason for this is that the input and output spaces of said parties are already occupied by the channels provided by~$\kket{\lambda_i}$.

We now demonstrate that~$W^k_\text{SQS}$ is a sufficient resource for winning the~$k$-cycle game with certainty.
The parties' strategies are as for the quantum switch above:
The sender implements the sender map~$S_s(x)$ in Eq.~\eqref{eq:sender_ex}, the receiver implements the receiver map~$R_{s^+}(a)$ in Eq.~\eqref{eq:receiver_ex}, all other parties do nothing, see Eq.~\eqref{eq:no_s no_r}.
Finally, $P$ activates the desired link by preparing
\begin{equation}
  M_P(s) = \projector{s}_{P_c} \otimes \frac{1}{2}\id_{P_t} \otimes \frac{1}{2^k}\id,
\end{equation}
where the maximally mixed state is prepared on the target system and on all dummy systems.
By contracting all maps, we obtain
\begin{equation}
  \begin{split}
    p(a|s,x)
    &=
    \frac{1}{2^{k+1}}
    \bbra{\lambda_s}\bbra{\Delta_s}
    \Big(
    \id \otimes S_s(x) \otimes R_{s^+}(a) \otimes
    \\
    &\bigotimes_{j\in[k]\setminus\{s,s^+\}} M_j^{\perp}
    \Big)
    \kket{\lambda_s}\kket{\Delta_s}
    =
    \delta_{xb}
    \,,
  \end{split}
\end{equation}
for all values of $s$, as desired.

Two remarks are in place.
First, note that in the above equation, the terms where the control system is~$\ketbra{i}{j}$ with~$i\neq j$ in~$W^k_\text{SQS}$ are irrelevant.
Hence, the same winning probability with the identical winning strategies are obtained for the special case of~$W^k_\text{SQS}$ with {\em classical control of causal order.}
More so, since the bit~$x$ is encoded and decoded in the same basis, the parties perfectly win the game using the same strategies on a purely {\em classical\/} process where {\em all off-diagonals\/} are canceled.
Second, the same calculations and the same reasoning hold for the {\em cyclic quantum switch.}

\section{Fixed-order and causal inequalities}
\label{sec:fixed and causal inequalities}
As shown above, the quantum switch and its simplified variants are sufficiently powerful to perfectly win the~$k$-cycle games.
The respective strategies require $P$, who acts in the global past of all other parties, to {\em adaptively\/} prepare the control system in the state~$\ket s$.
The value of~$s$, then again, is provided by the referee.
Crucially, this task does not require any quantum features, and in particular, no indefinite causal order.
The same is achieved with {\em classical control of causal order,} as opposed to quantum control.
In fact, as stated above, indefinite causal order in switch-like processes washes out if $F$'s outcome is ignored.
For the~$k$-cycle game this means that as long as~$F$ is non-playing, switch-like processes are effectively classical.

But then, under what circumstances may quantum-over-classical advantages arise with switch-like processes?
Our previous study hints at two central conditions.
First, party~$F$ must be a {\em playing\/} party and produce some outcome (at least in some cases).
Second, in order to suppress any power for establishing correlations stemming from the {\em dynamical\/} aspect, party~$P$ must act {\em non-adaptively.}
In such non-adaptive strategies, $P$ prepares the control system {\em independently\/} of~$s$, but possibly in a superposition, or entangled to other degrees of freedom.
With this, we effectively decouple the {\em dynamical\/} quality of switch-like processes and focus on the indefinite nature of the causal order.

Although for the present~$k$-cycle inequalities, quantum advantages are impossible, we analyze this non-adaptive case to further investigate the connection between dynamic and indefinite causal order.
Since~$P$ acts non-adaptively, we can absorb~$P$ into the process.
Moreover, since~$F$ is {\em irrelevant,} we can also absorb~$F$ into the process.
With this we obtain the setup as depicted in Fig.~\ref{fig:new2}:
All parties are {\em playing,} and we have~$k=n$.
Thus, in this setup, instead of referring to the~$k$-cycle game, we call it  the~$n$-cycle game\footnote{%
  A cycle through all parties is also known as a Hamiltonian cycle.
}.

In App.~\ref{app:The classical 2-switch},
we algebraically show that, using the classical two-party switch non-adaptively, it is {\em impossible\/} to violate the two-cycle inequality.
This unfeasibility in the classical case actually arises from the fact that a non-adaptive state preparation by $P$ produces but a probabilistic mixture of different causal order.

Remarkably, {\em any causal correlations\/} yield no advantage in winning the~$n$-cycle game when compared to fixed causal order.
Such ``causal'' observations can be formulated without referring to any underlying theory~\cite{ocb2012,oreshkov2016}.
To do so, consider --- as above --- the conditional probability distribution~$p(a|s,x)$, where~$a$ denotes the binary output of party~$s^+$,~$x$ is a binary input exclusively provided to party~$s$, and where all parties get as input~$s$.
These observations are {\em causal\/} whenever each party may only influence its future, including the causal order among the parties within its future.
Thus, the correlations~$p(a|s,x)$ are causal whenever it is possible to decompose them as
\begin{equation}
  \begin{split}
    p(a|s,x)
    =
    \sum_j
    p(j)
    \big(
      &
      [j=s^+]
      p_\text{indep}(a|s)
      \\
      &+
      [j\neq s^+]
      p_\text{dep}(a|s,x)
    \big)
    \,,
  \end{split}
  \label{eq:causaldef}
\end{equation}
where~$p(j)$ is a probability distribution over parties, indicating that party~$j$ acts {\em before\/} all other parties.
In the case where the first party ($j$) coincides with the guessing party~$s^+$ producing the outcome~$a$, then that party's output {\em cannot\/} depend on~$x$, as enforced with~$p_\text{indep}(a|s)$.
Otherwise, party~$j$ may influence the causal order of the remaining parties such that~$s^+$ is placed in the future of~$s$, and thus the output of the guessing party may depend on~$x$, as enabled via~$p_\text{dep}(a|s,x)$.
Since the value of~$s$ is uniformly distributed among~$n$ possibilities,
the guessing party acts before all others with probability~$1/n$.
Thus, the~$n$-cycle game is at best won with probability~$n^{-1}(1/2+(n-1))=1-\frac{1}{2n}$;
this bound {\em coincides\/} with the fixed-order bound presented above in Eq.~\eqref{eq:cycle_win_fixed}:
The~$n$-cycle inequality is {\em not only a fixed-order inequality, but also a causal inequality.}
Note that processes with quantum control of causal order are incapable of violating causal inequalities~\cite{oreshkov2016,araujo2015witnessing,ps2021,wechs2021}.
Therefore, any superposition of causal order among playing parties only cannot produce a violation the $k$-cycle inequality.

In Ref.~\cite{tselentis2023m}, some of us show that the~$k$-cycle inequality is facet-defining for the fixed-order correlations.
This means that the inequality is tight in the strongest sense.
In App.~\ref{app:facet}, we show that the transition from the~$k$-cycle inequality to the~$n$-cycle inequality preserves this property for {\em causal correlations:}
The~$n$-cycle inequality defines a facet of the set of causal correlations.

\section{Non-causal processes violating $k$-cycle inequalities}
\label{sec:Non-causal processes violating k-cycle inequalities}

As we show above, for non-adaptive strategies, $k$-cycle inequalities are effectively causal inequalities.
Hence, when $P$ acts non-adaptively, the $k$-cycle inequality can only be violated with processes that violate causal inequalities.
Examples of processes known to violate causal inequalities for any number of parties are presented in~\cite{baumeler2014maximal,araujo2017,svet2022,admissible2023}.
One of the simplest examples of these is the following three-party process~\cite{privateconv2014,baumeler2016space}
\begin{equation}
  \label{eq:w3}
    W_3 =\projector{w_3}
\end{equation}
with
\begin{align}
    &\ket{w_3}= \sum_{\substack{ijk\\\ell mn}} \ket{i,j,k}_P \ket{i\oplus \lnot m \land n}_{A^I} \ket{\ell}_{A^O}\\
    & \ket{j\oplus \lnot n \land \ell}_{B^I}\ket{m}_{B^O} \ket{k\oplus \lnot \ell \land m}_{C^I}\ket{n}_{C^O} \ket{\ell,m,n}_F
    \,, \nonumber
\end{align}
where all the Hilbert spaces involved are two-dimensional and $\oplus$ denotes the sum modulo two. The non-playing parties $P$ and $F$ emit and receive a three qubit state respectively. In order to violate the $k$-cycle inequality, the three playing parties have to collaborate in each round using, for example, the following operations
\begin{align}
    S_i(x)&=\mathds{1}_{I_i}\otimes\projector{x}_{O_i} \label{eq:senderL}\\
    R_i(a)&=\projector{a}_{I_i}\otimes \projector{0}_{O_i} \label{eq:receiverL}\\
    M_{i}^{\perp}&=\mathds{1}_{I_i}\otimes\projector{0}_{O_i}\,, \label{eq:thirdL}
\end{align}
where the operation $M^{\perp}$ of the third playing party now ensures that signaling according to the three-cycle $(ABC)$ is enabled by process $W_3$. For this strategy, we obtain the following probabilities for a general three-qubit state $\rho$ emitted by $P$
\begin{align}
    p(a|0,x) &= \bra{w_3}\rho_{P}\otimes S_A(x)\otimes R_{B}(a) \otimes M_{C}^{\perp} \otimes \mathds{1}_{F}\ket{w_3} \nonumber\\
    & =\sum_j \bra{j}\Tr_{02}(\rho) \ket{j}|\braket{j\oplus x|a}|^2 \\
    p(a|1,x) &= \bra{w_3}\rho_{P}\otimes M_{A}^{\perp} \otimes S_{B}(x) \otimes R_C(a) \otimes \mathds{1}_{F}\ket{w_3} \nonumber\\
    &=\sum_k \bra{k}\Tr_{01}(\rho) \ket{k}|\braket{k\oplus x|a}|^2 \\
    p(a|2,x) &= \bra{w_3}\rho_{P}\otimes R_{A}(a) \otimes  M_{B}^{\perp} \otimes S_C(x) \otimes \mathds{1}_{F}\ket{w_3} \nonumber\\
     &=\sum_i \bra{i}\Tr_{12}(\rho) \ket{i}|\braket{i\oplus x|a}|^2\,,
\end{align}
where we label the output Hilbert spaces of $P$ as $\mathcal{H}_0 \otimes \mathcal{H}_1 \otimes\mathcal{H}_2=\mathcal{H}_P$, and find that the three-cycle inequality is maximally violated whenever $P$ outputs $\projector{0}$ on the Hilbert space $\mathcal{H}_{s^+}$. Hence, we can first define the \emph{adaptive strategy}
\begin{equation}
\label{eq:dynamicL}
    M_P(s)=\sigma_{s,s^{-}}\otimes \projector{0}_{s^+}\,,
\end{equation}
where the $\sigma$'s are arbitrary two-qubit states, and obtain the winning probability
\begin{align}
  p^{\rm win}
  &=
  \frac{1}{6} \sum_x \Big(\Tr(\sigma_{0,2})+ \Tr(\sigma_{0,1})+\Tr(\sigma_{1,2})\Big)\delta_{xx}\\
  &=
  1\,,
\end{align}
for the three-cycle game. More importantly, however, for the process $W_3$ in Eq.~\eqref{eq:w3}, also the \emph{non-adaptive strategy}
\begin{equation}
\label{eq:fixedL}
    M_P= \projector{0,0,0}\,,
\end{equation}
gives a winning probability of one and, hence, a \emph{maximal violation} of the three-cycle inequality.
Note, however, that the process in Eq.~\eqref{eq:w3} is a \emph{classical} process beyond control of causal order~\cite{baumeler2014maximal}. Neither this process nor the above winning strategies have been discussed in Ref.~\cite{tselentis2023m}. Nevertheless, winning the~$k$-cycle game using general processes and non-adaptive strategies, again, does not represent a quantum-over-classical advantage.

\section{Conclusion and open questions}
We study the connection between the~$k$-cycle game, as introduced in Ref.~\cite{tselentis2023m}, and quantum processes with and without causal order. In that game, a referee challenges one of~$k$ players at random to communicate a bit to the cyclic successor of that player.
A winning chance of that game beyond~$1-\frac{1}{2k}$ proves the {\em incompatibility\/} of the correlations with {\em fixed causal relations\/} among the players.

It is known~\cite{tselentis2023m} that the~$k$-cycle game is perfectly won with purely classical switch-like processes that bear similarities to the {\em cyclic quantum switch.}
In the present work, we study the quantum switch and further introduce the {\em sparse quantum switch\/} --- a novel quantum process with minimal causal requirements ---, and show that they are sufficient for perfectly winning the~$k$-cycle game.
In the sparse quantum switch, the controlling party $P$, by preparing the control system in the state~$\ket s$, {\em activates\/} a communication link from~$s$ to its cyclic successor~$s^+$.
Communication to, from, and among all remaining parties is forbidden; they are causally disconnected.

We show that a violation of the~$k$-cycle inequality where~$k=n$, i.e., when the cycle traverses {\em all parties,} not only proves the incompatibility of the correlations with fixed causal relations,
but also their incompatibility with {\em dynamic causal relations.}
The resulting inequality, in fact, describes a {\em facet\/} of the causal polytope for any number of parties.
Simultaneously, it is known~\cite{oreshkov2016,araujo2015witnessing,ps2021,wechs2021} that {\em any\/} quantum process with quantum control is incapable of violating causal inequalities.
This provides a general proof of the insufficiency of such processes for violating the~$n$-cycle inequality.

Note that the~$n$-cycle game is similar to the ``guess your neighbor's input'' (GYNI) game~\cite{gyni2010} --- a central game in the field of non-locality.
In the GYNI game, all players must {\em simultaneously\/} communicate a random bit to their respective cyclic successor.
There, if the players are space-like separated, then a quantum-over-classical advantage is impossible.
However, there exists a non-signaling advantage for some input distributions.
Also, there is an advantage in winning that game with indefinite causal order when compared to any causal strategy~\cite{simplestcausalinequalities,liu2024},
and possibly a super-quantum over quantum advantage using indefinite causal order~\cite{boxworld2024}.

Finally, we show how processes ``without causal order'' can be invoked to perfectly win the~$n$-cycle game.
This proves that even though causal order must be abandoned, a violation of said inequality remains {\em in agreement with logical consistency.}

The present work raises a series of questions and opens several avenues.
It is at present unknown whether the {\em sparse\/} quantum switch provides any advantages in information processing.
Also, the {\em sparse\/} quantum switch can be straight-forwardly generalized to {\em any graph.}
Advantages in information-processing may then be related to graph-theoretic properties.
Moreover, the minimal causal requirements for that switch may be beneficial in extending the setups in Refs.~\cite{gppart3,tein2022,netdi2024,vanderlugt2024,chenwangwang2024} to any number of parties.
Another question that arises is whether a direct link can be drawn between the {\em degree of adaptivity\/} of~$P$'s strategy, e.g., in terms of distinguishability, and the degree of violation of the~$k$-cycle inequality.
Related to that, it might also be interesting to consider processes where more parties, and not only~$P$, may dynamically influence the causal order among other parties (see also the ``Grenoble'' process~\cite{wechs2021}),
or processes where discarding the future does not induce decoherence~\cite{wechs2021},
or even processes where the dynamic specification of the causal order is {\em not influenced\/} by any party~\cite{mothe2025}.
Also, the immediate similarly between the~$n$-cycle game and the GYNI game raises the question whether the established results for GYNI extend to our case:
Does there exist an input distribution that enables a non-signalling advantage for winning the~$n$-cycle game?

A core future avenue, for which we set the stage in this work, is the general question:
{\em Does a quantum-over-classical advantage emerge for the isolated quantum switch using non-adaptive strategies?}
In this paper we show that the $n$-cycle inequality is inadequate for showcasing such an advantage, if it exists. However, other candidate inequalities are outlined in Ref.~\cite{tselentis2023m}. Said inequalities, like the $k$-cycle inequality, concern a single input and output by the sender and receiver respectively. Generalized fixed-order inequalities with multiple inputs and outputs, which are a topic of active research and of independent interest, would provide further candidate inequalities. Could extensions of the $k$-cycle game, or any other game in Ref.~\cite{tselentis2023m}, to multiple inputs-outputs lead to advantages when indefinite causal order processes are used?
Recently, it has been shown~\cite{gppart3,tein2022,netdi2024,vanderlugt2024,chenwangwang2024} that under further non-signalling assumptions, quantum advantages become possible.
The setups described there, however, require additional parties to act space-like separated from all parties in the process.\footnote{%
  Can they be considered as certifications of the indefinite causal order present in the larger process comprising all parties, and where, in particular, the party in the global past acts non-adaptively?
}
Here, instead, we propose to seek a device-independent certification of indefinite causal order for processes with quantum control of causal order {\em in isolation\/} --- a term borrowed from Ref.~\cite{tein2022}.
For such a certification we regard as necessary to restrict party~$P$, situated in the past of everyone, to act {\em non-adaptively,}
i.e., to prepare the control system in a {\em constant\/} state.
That constraint removes any {\em dynamical aspect\/} that is also achievable with classical control, and allows for studying the essence of indefinite causal order.
Both possible answers to the above question have stark consequences.
A {\em negative\/} answer reproduces the central impossibility of violating causal inequalities~\cite{oreshkov2016,araujo2015witnessing,ps2021,wechs2021} on a ``lower'' level.
{\em This would indicate a core deficiency of quantum control of causal order.}
In contrast, a {\em positive\/} answer provides a {\em device-independent certification of the isolated quantum switch.}

\section*{Acknowledgments}
The authors thank \v{C}aslav Brukner for the initial discussion that started this project, and three anonymous referees for their valuable input. VB is funded by the Austrian Science Fund (FWF) grant No.\ ESP 520-N. \"AB is funded by the Swiss National Science Foundation (SNF) through project~214808. EET is supported by the Program of Concerted Research Actions (ARC) of the Universit\'e libre de Bruxelles and from the F.R.S.-FNRS under project CHEQS within the Excellence of Science (EOS) program.

\bibliography{references}
\newpage
\onecolumngrid

\appendix

\section{The classical two-switch}
\label{app:The classical 2-switch}

Here, we bound the winning probabilities of the two-cycle game with the classical switch process for both adaptive and non-adaptive strategies. The classical 2-switch is given by
\begin{equation}
   W^2_{\text{CS}}= \sum_{i}\projector{i}_{P_c}\otimes \kket{\pi_i}\bbra{\pi_i}\otimes \projector{i}_{F_c}
   \,,\label{eq:cswitch-2}
\end{equation}
where we only have two permutations of the playing parties:
\begin{align}
    \kket{\pi_0} &= \kket{\mathds{1}}_{P_t A^I} \otimes \kket{\mathds{1}}_{A^O B^I} \otimes \kket{\mathds{1}}_{B^O F_t} \,, \\
    &\text{and } \nonumber \\
    \kket{\pi_1} &= \kket{\mathds{1}}_{P_t B^I} \otimes \kket{\mathds{1}}_{B^O A^I} \otimes \kket{\mathds{1}}_{A^O F_t} \,.
\end{align}

Let us first consider the general \emph{non-adaptive strategy} given by Eq.~\eqref{eq:PFfixed_switch}, where the states $P$ emits is independent of the bit $s$ specifying the sender (and receiver). For general sender and receiver operations $S(x)$ and $R(a)$ we find that
\begin{align}
    p(a|0,x)
    &= \bra{0}\sigma\ket{0}\bbra{\pi_0}\rho_{P_t}\otimes S_A(x) \otimes R_B(a)\kket{\pi_0}   + \bra{1}\sigma\ket{1} \bbra{\pi_1}\rho_{P_t}\otimes S_A(x) \otimes R_B(a)\kket{\pi_1}\\
    &= p \sum_{\substack{ijlm}} \bra{l}\rho \ket{i}
    \bra{l,m}S(x)\ket{i,j} \bra{m}\Tr_O(R(a))\ket{j} +(1-p)\sum_{il} \bra{l}\rho \ket{i}
     \bra{l}\Tr_O(R(a))\ket{i} \,,
\end{align}
where we defined $p=\bra{0}\sigma\ket{0}$ and used that the sender map is assumed to be CPTP, and, therefore, $\sum_k \bra{m,k}S(x)\ket{j,k}= \bra{m}\mathds{1}\ket{j}=\delta_{mj}$ in the last equality. Analogously we have that
\begin{align}
  p(a|1,x)&=\bra{0}\sigma\ket{0}\bra{0}\bbra{\pi_0}\rho_{P_t}\otimes R_A(a) \otimes S_B(x)\kket{\pi_0} +\bra{1}\sigma\ket{1}\bbra{\pi_1}\rho_{P_t}\otimes R_A(a) \otimes S_B(x)\kket{\pi_1} \\
    &= p \sum_{il} \bra{l}\rho \ket{i} \bra{l}\Tr_O(R(a))\ket{i} +(1-p)\sum_{\substack{ijlm}} \bra{l}\rho \ket{i}
    \bra{m}\Tr_O(R(a))\ket{j} \bra{l,m}S(x)\ket{i,j} \,.
\end{align}
When calculating the winning probabilities, we need to sum the above expressions for $a=x$ and obtain
\begin{align}
    p^{\rm win}&= \frac{1}{4} \sum_x p(a=x|0,x)+ p(a=x|1,x) \\
               &=\frac{1}{4} \Big[ \Tr(\rho)+ \sum_{\substack{ijlm}} \bra{l}\rho \ket{i}  \cdot \sum_x
    \bra{m}\Tr_O(R(a))\ket{j} \bra{l,m}S(x)\ket{i,j}\Big] \,,
\end{align}
where we used that the receiver map, when summed over all possible guesses is assumed to be a CPTP map, i.e., $\sum_a \bra{m}\Tr_O(R(a))\ket{j}= \delta_{mj}$. Note that, the winning probability is independent of which control state $\sigma$ P emits.
To bound the above expression, we make use of the fact that $\Tr_{O}(R(a))$ is Choi matrix of a CP-map, i.e., a (sub)-normalized quantum state. For each $a$ we can write $\Tr_{O}(R(a))=\sum_m r_m(a) \projector{m}$ with $r_m \leq 1$ and chose to take the trace over the respective output space in the diagonal basis of $\Tr_{O}(R(a))$. Hence, we have
\begin{align}
\label{eq:bound1}
    \sum_{ijlm} \bra{l}\rho \ket{i}
    \bra{m}\Tr_O(R(a))\ket{j} \bra{l,m}S(x)\ket{i,j}
    &=
    \sum_{ijlm} \bra{l}\rho \ket{i} \bra{l,m} S(x)\ket{ij} r_m(a)\delta_{mj} \\
    &\le
    \sum_{ijlm} \bra{l}\rho \ket{i} \bra{l,m} S(x)\ket{ij}\delta_{mj} =\Tr(\rho)=1\,,
\end{align}
for all $a$. This means that the wining probability is bounded by
\begin{align}\label{eq: general fixed classical switch}
    p^{\rm win}\leq \frac{1}{4}(\Tr(\rho) + 2) \leq \frac{3}{4} \,,
\end{align}
which shows that the classical switch cannot violate the two-cycle inequality for non-adaptive strategies.\\

If, instead, we consider an \emph{adaptive strategy} given by Eq.~\eqref{eq:PFdynamic_switch1}, we find that
\begin{equation}
    p(a|0,x)= \bra{0}\sigma(0)\ket{0} \bbra{\pi_0}\rho_{P_t}\otimes S_A(x) \otimes R_B(a)\kket{\pi_0}
      + \bra{1}\sigma(0)\ket{1}\bbra{\pi_1}\rho_{P_t}\otimes S_A(x) \otimes R_B(a)\kket{\pi_1} \,,
\end{equation}
and
\begin{equation}
    p(a|1,x)=\bra{0}\sigma(1)\ket{0} \bbra{\pi_0}\rho_{P_t}\otimes R_A(a) \otimes S_B(x)\kket{\pi_0} +\bra{1}\sigma(1)\ket{1}\bbra{\pi_1}\rho_{P_t}\otimes R_A(a) \otimes S_B(x)\kket{\pi_1}
    \,,
\end{equation}
which gives the following bound for the winning probability:
\begin{equation}
    \label{eq:dynamicPwin}
    p^{\rm win}\leq \frac{1}{4} \left( 2+\bra{0}\sigma(0)\ket{0}+\bra{1}\sigma(1)\ket{1}\right) \,,
\end{equation}
where we again used the upper bound of Eq.~\eqref{eq:bound1}. Since $\bra{0}\sigma(0)\ket{0}+\bra{1}\sigma(1)\ket{1}$ is the sum of two probabilities, we have
\begin{equation}
     p^{\rm win}\leq 1  \,,
\end{equation}
where the equality is reached for $\sigma(0)=\projector{0}$ and $\sigma(1)=\projector{1}$, meaning that P emits the computational basis state of the control system according to bit $s$.

\section{Facet-defining causal inequalities}
\label{app:facet}
We study the cycle game with~$n$ playing parties (players)~\mbox{$\{0, 1, \dots, n-1\}$} and no non-playing parties.
In this scenario, a referee specifies one of the players as ``sender''~$s$, and announces this choice to every player.
Moreover, the referee provides a binary input~$x$ exclusively to player~$s$.
The player~$s^+=s+1\pmod n$ produces a binary output~$a$, and all other players do not produce an output.
Therefore, observations in this scenario correspond to a conditional probability distribution~$p(a|s,x)$.
Here, we abuse the notation slightly by using the same expression for the probability of output~$a$ given~$s$ and~$x$, and the set of such probabilities that constitute the probability distribution.
Such correlations are {\em causal\/} whenever the distribution admits a decomposition as in Eq.~\eqref{eq:causaldef}.

Geometrically, we represent each conditional probability distribution as a~$2n$-dimensional vector~$(p(0|s,x))_{s,x}$.
The polytope of causal correlations consists now of all such vectors that describe causal correlations.
First, note that this polytope is full-dimensional, i.e., the geometric representation is free of any redundancy.
The reason for this is that this polytope contains~$2n+1$ affinely independent vectors:
The all-zero vector (this corresponds to the strategy where~$s^+$ unconditionally outputs~$1$), and each unit vector.
To see that each unit vector is in that polytope, we consider, without loss of generality, the vector~$(1,0,\dots,0)$.
These correlations are produced by the following strategy.
First, the players agree to be arranged with player~$0$ first, i.e., the probability~$p(j=0)$ in Eq.~\eqref{eq:causaldef} is one.
Now, if~$s\neq 0$, then~$s^+$ outputs~$1$.
Alternatively, if~$s=0$, then~$s^+$ outputs~$x$.
This is possible because player~$0$, who is given~$x$, is in the causal past of player~$1$, who produces that output.

Now, we show that inequality in Eq.~\eqref{eq:cycle_win_fixed}, for~\mbox{$k=n$}, i.e.,
\begin{equation}
  \frac{1}{2n}\sum_{s,x}p(x|s,x) \leq 1-\frac{1}{2n}
  \,,
\end{equation}
describes a {\em facet,} i.e., a~$(2n-1)$-dimensional hyperplane, of that polytope.
We do this by identifying~$2n$ affinely independent vectors in the polytope that {\em saturate\/} that inequality.
First, we rewrite that inequality as
\begin{align}
  \sum_{s,x}p(x|s,x) &\leq 2n-1\quad \Leftrightarrow\\
  \sum_{s}\left(p(0|s,0)+(1-p(0|s,1)\right) &\leq 2n-1\quad \Leftrightarrow\\
  n+\sum_{s,x}(-1)^x p(0|s,x) &\leq 2n-1\quad \Leftrightarrow\\
  \sum_{s,x}(-1)^x p(0|s,x) &\leq n-1
  \,.
  \label{eq:ineqvect}
\end{align}
The first~$n$ of the vectors we are looking for are given by
\begin{equation}
  \{(1)_{s,x=0} || (\delta_{sk})_{s,x=1} \mid 0\leq k<n \}
  \,,
\end{equation}
where we use~$||$ to concatenate both~$n$-dimensional halves.
For~$k=0$, the following causal strategy yields the corresponding vector:
Player~$0$ is positioned first, causally followed by payer~$1$, etc.
If~$s^+=0$, then that player unconditionally outputs~$0$.
Otherwise, player~$s^+$ outputs~$x$.
These~$n$ vectors are affinely independent because each of them contributes to a {\em unique\/} dimension in the second half.
The remaining~$n$ vectors are analogous but where we simultaneously flip the entries at the coordinates~$(s=k,x=0)$ and~$(s=k,x=1)$:
\begin{equation}
  \{(1-\delta_{sk})_{s,x=0} || (0)_{s,x=1} \mid 0\leq k<n \}
  \,.
\end{equation}
The strategy is as above, where, for~$k=0$, player~$0$ unconditionally outputs~$1$ instead of~$0$.
These are also affinely independent because of their unique contribution in the first half.

\section{Proof of correctness for the sparse quantum switch}
\label{app:proof}
We prove that the sparse quantum switch given by Eq.~\eqref{eq:sqs} is a process matrix by an explicit representation of the process as a {\em quantum circuit with quantum control\/}~\cite{ps2021,wechs2021}.
The circuit is the following (read from left to right, draw using Quantikz~\cite{kay2018}):
\begin{equation}
  \begin{quantikz}
    &
    \wire[r][1]["\ket{s}"{above,pos=0},"c"{below,pos=0}]{a}&
    \ctrl{1}\gategroup[2,steps=3,style={dotted,rounded corners,fill=orange!15},background]{activated link}&
    \gate{\oplus}\wire[r][1]["\ket{s^+}"{above}]{a}&
    \ctrl{1}&
    &
    \gate{\oplus}\gategroup[3,steps=4,style={dotted,rounded corners,fill=blue!15},background]{nop}&
    \ctrl{1}&
    \ctrl{1}&
    \ctrl{1}&
    \ \ldots\ &
    \gate{\oplus}\gategroup[3,steps=4,style={dotted,rounded corners,fill=blue!15},background]{nop}&
    \ctrl{1}&
    \ctrl{1}\wire[r][1]["\ket{s^-}"{above}]{a}&
    \ctrl{1}&
\\
    &
    \wire[r][1]["t"{below,pos=0}]{a}&
    \gate[style={color=orange,draw=black}]{M_s}&
    &
    \gate[style={color=orange,draw=black}]{M_{s^+}}&
    &
    &
    \swap{1}&
    \gate[style={color=orange,draw=black}]{M_{s\oplus 2}}&
    \swap{1}&
    \ \ldots\ &
    &
    \swap{1}&
    \gate[style={color=orange,draw=black}]{M_{s^-}}&
    \swap{1}&
\\
    &
    \qwbundle{k}\wire[r][1]["\{D_i\}"{below,pos=0}]{a}&
    &
    &
    &
    &
    &
    \targX{}&
    &
    \targX{}&
    \ \ldots\ &
    &
    \targX{}&
    &
    \targX{}&
  \end{quantikz}
\end{equation}
The first two wires correspond to the ``control'' $c$ and to the ``target'' $t$ systems.
The control system is of dimension $k$ with the standard basis $\{\ket i\}_{i\in[k]}$, and the target system carries one qubit.
The last wire represents a bundle of~$k$ dummy systems~$\{D_i\}_{i\in[k]}$, each carrying one qubit.
The party in the past ($P$) initiates the states of all systems, and the party in the future ($F$) obtains the output.
In the circuit diagram, we moreover showcase the situation where~$P$ initiates the control system in the state~$\ket s$.
The gates labeled by~$\oplus$ increment the state on the control system by one modulo~$k$, i.e.,~$\ket{i} \mapsto \ket{i\oplus 1\pmod k}$.
The controlled swap gate exchanges the state of the target system~$t$ with the state of the dummy system~$D_i$, where~$i$ is the value of the control.
Finally, the colored controlled operations on the target system represent the operations of the parties: If the control is in the state~$\ket i$, then party~$i$ applies their operation~$M_i$.
These constitute the quantum control of the causal order.
The circuit is naturally understood in two parts.
In the first part (``activated link''), the target system prepared by~$P$ is processed by party~$s$ (also specified by~$P$), and thereafter by party~$s^+$.
In the second part (``nop''), dummy systems are forwarded to all remaining parties and immediately forwarded to the future~$F$.
The ``nop'' blocks in this second part are repeated~$k-2$ times.
Let us walk through the circuit, where~$P$ initiates the control in the state~$\ket s$.
In the first step, party~$s$ obtains the control system prepared by~$P$, and acts on it using~$M_s$.
Thereafter, the control system is incremented by one to the value~$s^+$.
Now, the party~$s^+$ obtains the target system that was previously processed by party~$s$.
So, in principle, party~$s$ may send a qubit to party~$s^+$.
This concludes the signaling part of the sparse quantum switch.
The following~$k-2$ repetitions of the ``nop'' blocks act as follows.
First, the state of the control system is incremented (modulo $k$).
Then, the dummy system of the party indicated by the control system is swapped with the target system.
This swap is undone at the end of each ``nop'' block.
Finally, that specified party interacts with the target system that now holds the dummy state.
We conclude that, independent of the state prepared by~$P$, each party acts once.
Also, at the end of the circuit the target system holds the initial states processed by~$s$ and thereafter by~$s^+$.
Finally, all other parties interact with the dummy system only and are incapable of communicating.
The~$(k-2)$-fold repletion of the ``nop'' block suggests that an implementation of the same circuit with lower depth may be possible.
We leave this possibility open.

\end{document}